# Classical and Quantum Theory of Photothermal Cavity Cooling of a Mechanical Oscillator

**Juan Restrepo[1], Julien Gabelli[2], Cristiano Ciuti[1], and Ivan Favero[1*]**

[1]*Laboratoire Matériaux et Phénomènes Quantiques, Université Paris Diderot, CNRS UMR 7162, 10 rue Alice Domon et Léonie Duquet, 75013 Paris, France*

[2] *Laboratoire de Physique des Solides, Université Paris-Sud, CNRS UMR 8502, Orsay, France.*

[*]*Corresponding author: ivan.favero@ univ-paris-diderot.fr*

Photothermal effects allow very efficient optomechanical coupling between mechanical degrees of freedom and photons. In the context of cavity cooling of a mechanical oscillator, the question of if the quantum ground state of the oscillator can be reached using photothermal back-action has been debated and remains an open question. Here we address this problem by complementary classical and quantum calculations. Both lead us to conclude that: first, the ground-state can indeed be reached using photothermal cavity cooling, second, it can be reached in a regime where the cavity detuning is small allowing a large amount of photons to enter the cavity.

# 1. Introduction

Optomechanical systems have recently made impressive progress with applications ranging from sensing to fundamental tests of quantum mechanics [1-3]. After first quantum control experiments performed on a 6 GHz thin plate oscillator [4], one of the challenges in the field of optomechanics is now to allow other mechanical oscillators of smaller frequency and larger mass to also enter the quantum regime. This can in principle be achieved by means of cavity self-cooling and recent work combining cavity cooling and conventional cryogenics have indeed shown progress in this direction [5-9]. The optomechanical coupling generally relies on radiation pressure in these experiments and the "good cavity" condition, where the cavity photon lifetime exceeds the mechanical oscillator period, must be fulfilled to hope obtaining phonon occupation number of the oscillator below one [10-13]. However in this regime, the pump field is far detuned from the cavity resonance, restricting the amount of photons injected in the cavity and hence the magnitude of the optical cooling mechanism.

Actually the first proof-of-principle experiment of cavity self-cooling of a mechanical oscillator was performed using photothermal optomechanical coupling rather than a mere radiation pressure coupling [14]. In photothermal effects (also sometimes named bolometric effects), photons are absorbed by the mechanical oscillator and give birth to a thermo-elastic distortion, which displaces the oscillator. One advantage of photothermal pressure is that it can be orders of magnitude larger than radiation pressure. Indeed in the process of radiation pressure a photon is reflected on a moving mirror of velocity v and its energy E is shifted by an amount of orders (v/c)E by the Doppler effect [15]. The mechanical energy given by the photon to the mirror during this process is extremely small. In contrast, in the case of photothermal pressure, the photon is absorbed and transfers its whole energy E to the mirror. This can result in a large effective optical force acting on the mirror, provided that the mirror thermo-elastic properties are optimized. As detailed in [16], in experimental situations the photothermal force can easily overcome radiation pressure force by several orders of magnitude and can even have its direction opposed with respect to radiation pressure. These properties have led to original situations of simultaneous cavity cooling of several mechanical modes [16-17] or to the study of rich non-linear cavity dynamics driven by bolometric forces [18].

Still, for what concerns cavity self-cooling of the mechanical oscillator to its quantum ground-state, photothermal coupling has been comparatively very little investigated [14,19-20]. To our knowledge, a complete discussion of quantum limits of photothermal cavity cooling is still lacking. This results from the fact that being non-conservative in nature, the photothermal interaction is difficult to describe with a quantum Hamiltonian, in contrast to the cavity radiation pressure case where a solid Hamiltonian is available for quantum optics calculations [21]. The discussion of these limits is however important given the cooling efficiency observed in photothermal cavity self-cooling experiments [14]. Two main features distinguish photothermal pressure from radiation pressure in this context: first, its very large relative amplitude, which can produce similar effects for smaller optical intensity, second its distinct dynamical behavior and typical timescale, which result in a qualitatively different noise spectrum. In this article, we aim at treating the problem of photothermal cavity cooling of a mechanical oscillator, trying to understand how we can benefit from these two distinct features. For simplicity, we will assume the photothermal pressure to act on the oscillator in the same direction as radiation pressure.

The first section deals with a classical approach to the problem, where we draw the conclusion that photothermal cavity-cooling of the oscillator to its quantum ground-state is feasible even in the "bad cavity" limit and for a moderate detuning of the pump field to the cavity, allowing many photons to enter the cavity. In the second section, this conclusion is confirmed by a more general quantum approach where the full quantum noise spectrum of the force is derived. Interestingly, this quantum approach allows understanding the classical expression of cooling as resulting from an interference between radiation pressure and photothermal effects.

## 2. Classical approach

There is a lot to learn first from a classical treatment of the problem, before assessing quantum limits more rigorously. Indeed when a quantum harmonic oscillator of angular frequency $\omega_0$ is coupled linearly to a bath of oscillators, its variance fluctuations can be computed in a classical manner, provided that the bath temperature is large with respect to $\hbar\omega_0$ and provided that zero-point fluctuations of the oscillator are added "by hands" at the end of the calculation. This interesting property relies on the harmonicity and linearity of the coupling to the bath. The property appears for example obviously in the standard quantum Langevin description of a damped harmonic oscillator [22]. If we restrict ourselves to a linearized approach, we can hence draw conclusions about quantum limits of photothermal cooling from a purely classical calculation.

The mechanical oscillator is considered to be harmonic here and its dynamical equation in absence of light is simply given by the Newton equation:

$m\ddot{x} + m\Gamma\dot{x} + Kx = F_{Langevin}$

K and m are the spring constant and mass of the oscillator. The mechanical damping parameter $\Gamma$ relates to the classical Langevin force $F_{Langevin}$ by usual fluctuation-dissipation relation. $\Gamma$

contains contributions from different loss sources amongst which thermoelastic damping is of special interest for our discussion. Indeed, like the photothermal force, the Langevin force associated to thermoelastic damping relies on the thermal expansion of the material. Under illumination of the oscillator, it can hence be seen as a fluctuating component of the photothermal force, finding its origin in the steady state temperature fluctuations of the oscillator body. In our description, this contribution will be included in $\Gamma$ and $F_{Langevin}$ and not further detailed. We will on the contrary focus on the fluctuations of the photothermal force originating from fluctuations in the number of photons absorbed by the oscillator. This second contribution was already discussed in the context of gravitational waves interferometers in ref [23-24] and dubbed "photothermal shot noise". We aim here at understanding how photothermal shot-noise interplays with photothermal cavity self-cooling of a mechanical oscillator.

The oscillator is from now on the movable back-mirror of a Fabry-Pérot cavity whose lossless front mirror has transmission T. The oscillator dynamical equation is given by:

$$m\ddot{x}(t) + m\Gamma\dot{x}(t) + Kx(t) = F_{Langevin}(t) + F_{photothermal}(t) \quad (1)$$

$$F_{photothermal}(t) = \beta \frac{2R}{c} \int_{-\infty}^{+\infty} h(t-u) P_{abs}(u) du$$

where $P_{abs}(t) = AP_{circ}(t)$ with $P_{circ}(t)$ the circulating power in the cavity and $P_{abs}(t)$ the power absorbed by the oscillating mirror. A is the absorption coefficient of the movable mirror, R its reflectivity and its transmission is taken to be zero. The h function accounts for a general linear response of the photothermal force upon absorption of photons by the oscillator. The function h must respect the causality principle and reflect a thermal relaxation process with timescale $\tau_{th}$ hence we set $h(t)=(1/\tau_{th})\Theta(t)\exp(-t/\tau_{th})$ with $\Theta$ a Heaviside function. The Fourier transform of h is $h(\omega)=1/(1+i\omega\tau_{th})$ with the following definition:

$$f(\omega) = \int_{-\infty}^{+\infty} f(t)e^{-i\omega t}\, dt \qquad (2)$$

The expression of the photothermal force chosen in Eq. 1 is equivalent to that used in ref [25] and is transformed to that of ref [16] by a simple integration by parts. In Eq. 1, the photothermal force is expressed in units of the radiation pressure force acting on the mirror $F_{rad}=(2R/c)P_{circ}$ in the sense that for a constant illumination and for a rigid cavity $F_{photothermal}=\beta A F_{rad}$. The discussion is focused here on cases where photothermal pressure overcomes radiation pressure by far hence we will consider $\beta A \gg 1$ and neglect the radiation pressure force in the classical dynamics of the oscillating mirror. As we will see in the next section, the inclusion of radiation pressure in the calculation does not alter the main conclusions reached in this section.

When the movable mirror now oscillates, the cavity length is modulated around its average value $L_0$ and the mirror motion $x(t)$ couples to the light power circulating in the cavity through Eq. 1. Conversely the steady-state average circulating power $P_{circ}$ couples to the mirror coordinate x through the Fabry-Pérot response function.

$$P_{circ}(x) = \frac{T/\tau_0^2}{\kappa^2 + (-\Delta + 2\omega_L x/L_0)^2} P_{inc} \qquad (3)$$

where $\Delta=\omega_c-\omega_L$ is the detuning of the laser to the cavity resonance of length $L_0$, $\tau_0 = 2L_0/c$ is the corresponding cavity round-trip time, $\kappa=(T+A)/2\tau_0$ is the cavity field decay rate and $P_{inc}$ the incident power on the cavity. The mutual coupling expressed by differential equations 1 and 3 leads to a rich non-linear photothermal dynamics of the cavity-mirror system that was already explored theoretically and experimentally, for example in [18]. Here we are focusing on the position variance of the oscillator under self-cooling hence we will restrict ourselves to linearizing these two equations for small oscillator displacements |x(t)|<< $L_0$. The average absorbed power becomes:

$$P_{abs}(x) = P_{abs}(0) + x.(dP_{abs}/dx)_{x=0} \qquad (4)$$
$$= (1 + 4x\frac{\Delta \omega_L}{L_0(\kappa^2 + \Delta^2)})\frac{TA/\tau_0^2}{\kappa^2 + \Delta^2}P_{inc}$$

For example, if the laser line sits on a flank on the cavity resonance ($\Delta=\kappa$) and if A=T, the gradient of absorbed power upon mirror motion is $dP_{abs}/dx = (8F/\lambda)P_{abs}$, where $F=2\pi/(T+A)$ is the cavity finesse and $\lambda$ is the laser wavelength. We will assume here that the cavity response time is small compared to $\tau_{th}$, a condition which is usually fulfilled in experiments. This condition amounts to saying that $P_{abs}(u)$ in Eq.1 can be replaced by the sum of the average power $P_{abs}(x(u))$ and a fluctuating term $\delta P_{abs}(u)$. The static term $P_{abs}(0)$ produces a constant photothermal force which shifts the equilibrium position of the oscillator. We will omit this shift in what follows by considering fluctuations x around the new equilibrium position. Transforming Eq. 1 into Fourier space we obtain:

$$-m\omega^2 x(\omega) + im\Gamma\omega x(\omega) + Kx(\omega) = F_{Langevin}(\omega) + \beta\frac{2R}{c}h(\omega)\delta P_{abs}(\omega) + \beta\frac{2R}{c}(dP_{abs}/dx)h(\omega)x(\omega) \qquad (5)$$

which can be recast in the more compact form:

$$\left[\omega_{eff}^2 - \omega^2 + i\Gamma_{eff}\omega\right]x(\omega) = \frac{1}{m}\left[F_{Langevin}(\omega) + \beta\frac{2R}{c}h(\omega)\delta P_{abs}(\omega)\right] \qquad (6)$$

where we have introduced the effective eigenfrequency $\omega_{eff}$ and damping $\Gamma_{eff}$ of the oscillator under photothermal back-action induced by the cavity :

$$\omega_{eff}^2 = \omega_0^2 (1 - \beta \frac{1}{1+\omega_0^2 \tau_{th}^2} \frac{2R}{c} (dP_{abs}/dx) \frac{1}{K}) \qquad (7)$$

$$\Gamma_{eff} = \Gamma(1 + Q_m \beta \frac{\omega_0 \tau_{th}}{1+\omega_0^2 \tau_{th}^2} \frac{2R}{c} (dP_{abs}/dx) \frac{1}{K})$$

with $\omega_0^2$=K/m the bare mechanical oscillator angular frequency and $Q_m$=$\omega_0$/$\Gamma$. These approximate expressions were already derived in [14,18]. The important difference now is that we have included fluctuations of the photothermal force in Eq. 6 and most importantly the photothermal shot-noise represented by δP. These fluctuations play an important role since they generally counteract the cavity cooling mechanism. In case of cavity self-cooling by radiation pressure for example, it is necessary to operate in the good cavity limit ($\omega_0$>κ) to avoid that these fluctuations preclude reaching the oscillator quantum ground state, as discussed by several authors [10-13].

In a Fabry-Pérot cavity, it is known that fluctuations of the circulating power have a super-Poissonian statistics. For a coherent state at the cavity input, an extension of calculations of ref [24] leads to the following (two-sided) noise power spectral density:

$$S_{P_{circ}}(\omega) = \frac{F}{\pi} \hbar \omega_L P_{circ} (\frac{1}{1+(\frac{\omega - \Delta}{\kappa})^2} + \frac{1}{1+(\frac{\omega + \Delta}{\kappa})^2}) \qquad (8)$$

with the relation δ(ω+ω')$S_{Pcirc}$=(1/2π)<$P_{circ}$(ω)$P_{circ}$(ω')> and $P_{circ}$ the average circulating power. One might think at first that these fluctuations would transfer to a super-Poissonian statistics of the power absorbed by the movable back-mirror but this is not the case. Indeed the absorption process can be viewed from the photons point of view as the introduction of a beam-splitter of transmission A at the cavity output port. The situation is hence analog to two lossless mirrors forming a symmetric Fabry-Pérot cavity. In this case the transmitted photons have Poissonian statistics for a coherent state input. For a Poissonian input beam, the statistics of absorbed photons in our lossy cavity is hence Poissonian and not super-Poissonian. This statistical effect results from the coupling of cavity photons to vacuum fluctuations during the absorption process [24].

$$S_{\delta P_{abs}}(\omega) = A\hbar\omega_L P_{circ} = \hbar\omega_L P_{abs} \qquad (9)$$

where $P_{abs}=AP_{circ}$ is the average absorbed power. This white approximation of the noise is excellent for the situation of interest here where the mechanical frequency is several orders of magnitude smaller than the laser frequency. We obtain from Eq. 6:

$$|x(\omega)|^2 = \frac{1}{m^2}\frac{1}{(\omega_{eff}^2-\omega^2)^2+(\Gamma_{eff}\omega)^2}\left[|F_{Langevin}(\omega)|^2 + 2\pi\beta^2\left(\frac{2R}{c}\right)^2\frac{1}{1+\omega^2\tau_{th}^2}\hbar\omega_L P_{abs}(x=0)\right] \qquad (10)$$

where the (two-sided) spectral density of the equilibrium Langevin force is $(1/2\pi)|F_{Langevin}(\omega)|^2$ $=2k_B Tm\Gamma$ [16] with T the temperature of the oscillator environment ($k_B T \gg \hbar\omega_0$). If we assume the oscillator relaxation time to be large compared to the correlation time of the driving force, the fluctuation $<x^2>_{cl}=(1/2\pi)^2\int<x(\omega)x(-\omega)>d\omega$ is determined by the noise force spectrum at $\pm\omega_{eff}$. We obtain the classical part of the oscillator fluctuations:

$$m\omega_{eff}^2\langle x^2\rangle_{cl} = \frac{1}{\Gamma_{eff}}\left[\Gamma k_B T + \frac{1}{2m}\beta^2\left(\frac{2R}{c}\right)^2\frac{1}{1+\omega_{eff}^2\tau_{th}^2}A\hbar\omega_L P_{circ}(x=0)\right] \qquad (11)$$

At this stage of the derivation, the zero-point fluctuations of the mechanical oscillator are absent. As mentioned above, they can now be included by hands to obtain the correct quantum result $m\omega_{eff}^2<x^2>=m\omega_{eff}^2<x^2>_{cl}+\hbar\omega_{eff}/2$. We can define an effective fluctuation temperature by $k_B T_{eff}$

= $m\omega_{eff}^2 \langle x^2 \rangle$. With these notations, the quantum ground-state of the oscillator is reached when $k_B T_{eff}$ becomes of the order of $\hbar\omega_{eff}/2$.

Amongst the different regimes expected from Eq. 11, we are especially interested in the advantageous case of strong cooling, where $\Gamma_{eff} \gg \Gamma$ and $\omega_{eff} \sim \omega_0$. In this regime, the thermal fluctuations are damped to a negligible amount. As discussed in several works [10,16], we remind that reaching this regime requires having a sufficiently large mechanical quality factor to start with in the experiments. In practice $Q_m$ must be larger than the initial average population of the oscillator. In this strong cooling regime the oscillator fluctuations are determined by the photothermal shot-noise:

$$K_{eff} \langle x^2 \rangle_{cl} = \beta R \hbar \frac{1}{\tau_{th}} \frac{\tau_0(\kappa^2 + \Delta^2)}{8\Delta} \qquad (12)$$

Let us consider a first simple situation where the laser line is red-detuned by half of the cavity resonance width $\Delta = \kappa$ with $R \sim 1$ and $A \gg T$. In this case the cavity losses are dominated by absorption and we obtain:

$$K_{eff} \langle x^2 \rangle_{cl} \sim \frac{1}{8} \beta A \frac{\hbar}{\tau_{th}} \qquad (13)$$

which can be made much smaller than $\hbar\omega_0/2$ provided that $\omega_0 \tau_{th} \gg (1/4)\beta A$. This means that photothermal cavity cooling to the ground-state is feasible provided that the thermal time lag is large enough compared to $\beta A/\omega_0$. This regime is reminiscent of the good cavity limit of radiation pressure cavity cooling, with the role of the cavity photons lifetime now played by the thermal relaxation time. Indeed in this regime the cavity photothermal back-action efficiently damps Brownian fluctuations of the oscillator, but with a limited amount of added force fluctuations thanks to a low-pass "thermal low-pass frequency filter" present in the force noise.

Fig. 1a shows the noise spectral density of the photothermal force normalized to the maximal noise spectral density of radiation pressure force, which is obtained when $\omega=\Delta$ (see Eq. 8). In Fig. 1a we have $\beta A=100$. At low frequency, the fluctuations of the photothermal force are very important, reflecting the fact that photothermal pressure is itself amplified by a factor $\beta A$ with respect to radiation pressure. In this range, the photothermal shot-noise precludes reaching the ground state because of added fluctuations. These fluctuations vanish when raising the frequency $\omega_0$.

For $\omega_0\tau_{th} \sim \beta A$, the photothermal force shot-noise driving the oscillator becomes inferior to radiation pressure shot-noise level. A the same time, the photothermal back-action damping the oscillator is typically still a factor $\beta A$ more efficient than radiation pressure damping. In this regime and for values of $\omega_0\tau_{th}$ ~in excess of $\beta A$, the ground-state can be approached even in the bad-cavity limit where $\omega_0/\kappa$ is inferior to one. Fig. 1b shows a two-dimensional plot of $m\omega_0^2<x^2>/(\hbar\omega_0/2)$ in the strong cooling regime as a function of $\omega_0\tau_{th}$ and $\Delta/\kappa$ for the same numerical parameters as in Fig. 1a and for $\tau_{th}=1$ms. This plot is obtained from Eq. 12. As discussed above, if the condition $\omega_0\tau_{th} >> \beta A$ is fulfilled, we observe that the amount of classical fluctuations can be made smaller than quantum fluctuations for detuning typically superior or equal to 1. This leads to a total normalized fluctuation temperature approaching 1. At smaller detuning on the contrary, the cavity cooling mechanism is too un-efficient to approach the ground-state.

Interestingly, in contrast to the radiation pressure case where the laser needs to be far-detuned to the cavity resonance, this original "photothermal ground-state cooling regime" can here be obtained close to cavity resonance (Δ~κ), ensuring a large amount of photons in the cavity, which could prove extremely useful in the experiments. In radiation pressure cavity cooling experiments, the small amount of useful photons in the cavity in the good-cavity limit is a severe limitation.

In the next section, we will confirm these conclusions by a more general quantum calculation where both radiation pressure and photothermal pressure are included. If quantitative aspects are modified in some regimes, the competition between cavity self-cooling and added fluctuations is qualitatively un-altered provided that βA is large compared to 1.

Before closing this section, we note that the classical calculation shown here could be extended to the case of radiation pressure. In the case where A=0, photothermal effects vanish and our classical calculation leads the standard result that ground-state cooling is only attainable in the good-cavity limit. As discussed above, no quantum theory is needed to reach this conclusion: the calculation just needs to include classically the effect of shot-noise and zero-point fluctuation of the oscillator to obtain the correct quantum result (at least in the harmonic and linearized case).

## 3. Quantum approach

We will now treat the problem more rigorously within the quantum formalism. A first remark is that, due to the dissipative nature of the photothermal interaction, it is difficult to derive the Hamiltonian of the corresponding optomechanical coupling. In the photothermal effect indeed photons are absorbed. The absorbed energy is then distributed over an infinite number of degrees of freedom, giving rise to a temperature increase, which produces a displacement of the oscillator. In the case of radiation pressure in contrast, the conservative nature of the force allows the derivation of the Hamiltonian, and the radiation pressure Hamiltonian for an absorption-free cavity has been indeed known for a long time [21]. This Hamiltonian has served as a base ingredient for many theoretical works relating to radiation pressure effects in cavities.

Nevertheless, even if the system consisting of the cavity field and the mechanical oscillator does not conserve its energy when coupled through photothermal pressure, it is still possible to write down a Hamiltonian formulation of the mechanical oscillator dynamical evolution alone. The same holds true generally in the case of a quantum mechanical system driven by a (non-conservative) classical force [26]. The quantum evolution of the system is in this case obtained by including a classical forcing -**x**f(t) in the mechanical oscillator Hamiltonian **H**=$\hbar\omega_0$(**b**$^+$**b**+1/2)-**x**f(t) where **b** is the oscillator annihilation operator and **x** the oscillator position operator. In this case, as in time-dependant perturbation theory in quantum mechanics, the Hamiltonian directly depends on time, reflecting an energy flow between the oscillator and its environment. But the Heisenberg equations still lead the correct dynamical evolution of the oscillator operators. Here we will use this approach with an expression of the photothermal force f(t) reminiscent of Eq. 1, where a thermal response function h(t) is convoluted with the operator number of photons absorbed by the oscillator **n**$_{abs}$. In this approach quantum fluctuations are accounted for using commutations relations of **x** for the oscillator and **n**$_{abs}$ for photons. The classical part of the dynamics enters the description through the linear response h(t).

An interesting article [25] already employed such approach to discuss the limits of radiation pressure and photothermal cavity cooling. In this article, the photothermal force amplitude was considered to be at most half of that of the radiation pressure force, with the underlying idea that an absorbed photon transmits $\hbar k$ momentum to the mirror upon absorption in comparison to $2\hbar k$ when the photon is reflected. The conclusions reached in [25] were hence limited to the very special case where both pressures are comparable in magnitude. In reality, as already discussed in the introduction, the photothermal pressure can be orders of magnitude larger than radiation pressure for a given incident power on the mirror and it is actually this interesting property that we want to use in the present context.

Here we will follow closely the path of ref [25] and redo the calculations in the case of a dominating photothermal interaction. These calculations are slightly more general than that of section 1, since they now also include radiation pressure. Interestingly, they allow us to recover the outcome of the classical calculations and to give additional insights into the conclusions reached in section 1.

The Heisenberg equations allow to write a kind of "Newton law" for the oscillator position operator **x**.

$$m\ddot{x} + m\Gamma\dot{x} + Kx = F_{Langevin}(t) + F_{opt}(t) \qquad (14)$$

$$F_{opt}(t) = \frac{\sqrt{2}\hbar k}{\tau_0} I(t) + \beta\sqrt{2}\hbar k \int_{-\infty}^{t} \frac{du}{\tau_{th}} e^{-(\frac{t-u}{\tau_{th}})} I_{abs}(u)$$

with the same notations as in the first section, and with a close match with those of ref [25]. Now $I(t)=a^{+}(t)a(t)$ is the intracavity intensity operator with $a(t)$ the intracavity field annihilation operator. The intensity absorbed by the movable back-mirror is represented by $I_{abs}(t)=a^{+}_{abs}(t)a_{abs}(t)$ where $a_{abs}(t)=\sqrt{(A/\tau_0)}a(t)-b_{in}(t)$ and $b_{in}(t)$ is the annihilation operator corresponding to vacuum fluctuations entering the cavity through the absorption process [24]. Note that with this definition, $I(t)$ and $I_{abs}(t)$ do not have the same dimension. The cavity field dynamical equation is given by:

$$\dot{\mathbf{a}}(t) = -(\kappa + i\Delta_c)\mathbf{a}(t) + i\frac{k}{\tau_0\sqrt{2}}\mathbf{x}(t)\mathbf{a}(t) + \sqrt{T/\tau_0}\mathbf{a}_{in}(t) + \sqrt{A/\tau_0}\mathbf{b}_{in}(t) \qquad (15)$$

with $\mathbf{a}_{in}(t)$ the annihilation operator of the field injected in the cavity through the front-mirror, $\Delta_c$ the empty cavity detuning and k=2π/λ. The advantage of the approach used here, which models the absorption process as "effective mirror" of transmission A, is that it allows to directly recover the correct Poissonian statistics of absorbed photons, since all vacuum fluctuations are now automatically accounted for.

An important difference with respect to ref [25] is that the parameter β can here be much larger than one, reflecting the possibility to have a photothermal pressure overcoming radiation pressure by far. As mentioned in the introduction, β could even be negative. We will however restrict ourselves to the positive case for clarity.

The semi-classical steady state mean value of the position operator now reads:

$$\langle x \rangle = \frac{\sqrt{2}\hbar k}{K\tau_0}(1+\beta A)\alpha^2 \qquad (16)$$

with α=<a> the mean value of operator **a** and

$$\Delta_{nl} = \frac{\hbar k^2}{K\tau_0^2}(1+\beta A)\alpha^2 \qquad (17)$$

the non-linear detuning induced by the steady-state intracavity photons pressure on the movable mirror. The overall cavity detuning Δ is the sum of the empty cavity detuning $\Delta_c$ and of this non-linear term. $\Delta=\Delta_c-\Delta_{nl}$.

To study the dynamics of the system and its quantum fluctuations, we proceed linearizing the operators around these semi-classical mean-values **x**=<**x**>+δ**x**, **a**=α+δ**a**, **a**$_{in}$=<**a**$_{in}$>+δ**a**$_{in}$ and **b**$_{in}$=0+δ**b**$_{in}$ (no classical drive at the cavity output port). The linearized optical force fluctuation contains a part proportional to δ**x** as a result of the optomechanical coupling. The rest relates to fluctuations of the input light source and is named thereafter δ**f**$_{opt}$.

Equations (14) and (15) are Fourier transformed and lead to first order in the fluctuations to:

$$\delta x(\omega) = \chi_{eff}(\omega)\left[\delta F_{Langevin}(\omega) + \delta f_{opt}(\omega)\right] \quad (18)$$

with

$$\chi_{eff}^{-1}(\omega) = m(\omega_0^2 - \omega^2 + i\Gamma\omega) - 2\Delta\frac{\hbar k^2}{\tau_0^2}\alpha^2(1 + \frac{\beta A}{1 + i\omega\tau_{th}})\left[\frac{1}{D(\omega)D^*(-\omega)}\right] \quad (19)$$

with $D(\omega) = \kappa + i\Delta + i\omega$. Eq. 18 and 19 correspond to Eq. 6 in the classical calculation of the first section. Now $\chi_{eff}(\omega)$ is a complex number and **x**, **F**$_{Langevin}$ and **f**$_{opt}$ are operators.

We find back Eq. 15 of ref [25] relating to δ**f**$_{opt}$ with the only obvious change that A has to be replaced by βA each time A appears in connection with the "thermal filter term" $1/(1+i\omega\tau_{th})$.

We proceed with the calculations of the variances and associated quantum noise spectra. The quantum noise spectrum corresponding to an operator **x** is defined via

$$\langle \delta x(\omega) \delta x(\omega')\rangle = 2\pi\delta(\omega + \omega')S_x(\omega) \quad (20)$$

leading for the position variance:

$$\Delta x^2 = \int_{-\infty}^{+\infty} S_x(\omega)\frac{d\omega}{2\pi} \quad (21)$$

with from Eq. 18:

$$S_x(\omega) = |\chi_{eff}(\omega)|^2 (S_{F_{Langevin}}(\omega) + S_{f_{opt}}(\omega)) \quad (22)$$

For clarity, we adopt the same normalized notation as in ref [25]: $b=\omega_0/\kappa$ the hybrid quality factor, $\varphi=\Delta/\kappa$ the normalized detuning, $\varphi_{nl}=\Delta_{nl}/\kappa(1+\beta A)$, $d=\omega_0\tau_{th}$ and $\Omega=\omega/\omega_0$. We also normalize the position operator **x** to the zero-point fluctuation of the oscillator by defining a normalized $\mathbf{X}=\mathbf{x}\sqrt{(m\omega_0/\hbar)}$ and find the following equation:

$$\Delta X^2 = \int_{-\infty}^{+\infty} \frac{d\Omega}{2\pi} \frac{1}{(1-\Omega^2+\delta\Omega)^2 + (\Omega/Q+\delta\Gamma)^2}\left[\widehat{S}_{F_{Langevin}}(\Omega) + \varphi_{nl}\widehat{S}_{f_{opt}}(\Omega)\right] \qquad (23)$$

with the appearance of normalized noise spectra both for the Langevin and optical force. Note that, instead of being single-sided like in ref [25], the integral is now taken from $-\infty$ to $+\infty$ to account for a possible angular frequency asymmetry in the noise spectrum. In Eq. 23, $\delta\omega$ and $\delta\Gamma$ correspond to a resonance shift (real part of the susceptibility) and an optically modified damping (imaginary part of the susceptibility).

We found the normalized noise spectrum of the optical force (including both radiation pressure and photothermal pressure) to be:

$$\widehat{S}_{f_{opt}} = \frac{1}{(1-b^2\Omega^2+\varphi^2)^2+(4b^2\Omega^2)}\left\{\begin{array}{l}\frac{2T}{T+A}\left|1+\beta\frac{A}{1+i\Omega d}\right|^2(1+\varphi^2+b^2\Omega^2-2b\Omega\varphi)+\\ \frac{2A}{T+A}\left|(1+i\Omega b-i\varphi)\left[1+\beta\frac{T+A}{2(1+i\Omega d)}\left(\frac{A-T}{T+A}-i\varphi-ib\Omega\right)\right]\right|^2\end{array}\right\} \qquad (24)$$

Eq. 24 is different from the result found in [25], with the important difference that the spectrum is now asymmetric in frequency, reflecting the asymmetry between emission and absorption process. In contrast to refs [10,27], our choice of convention (Eq. 2) implies that $S_{fopt}(+\omega)$ corresponds to the emission of energy quanta by the cavity into the mechanical oscillator.

We will focus on the case where the effective damping time of the mechanical oscillator remains large in comparison to the correlation time of optical and Langevin forces. This means that the photon-dressed mechanical oscillator has a peaked frequency response. In this limit the frequency shift δω and effective damping $\Gamma_{eff}$ are given by:

$$\delta\omega = \frac{-2\varphi\varphi_{nl}}{(1-b^2+\varphi^2)^2+4b^2}((1-b^2+\varphi^2)(1+\frac{\beta A}{1+d^2})-\frac{2\beta Abd}{1+d^2}) \qquad (25)$$

$$\Gamma_{eff} = \Gamma\left[1+\frac{2\varphi\varphi_{nl}Q}{(1-b^2+\varphi^2)^2+4b^2}((1-b^2+\varphi^2)\frac{\beta Ad}{1+d^2}+2b(1+\frac{\beta A}{1+d^2}))\right] \qquad (26)$$

These expressions correspond to Eq. 7 obtained by classical calculations in the first section. Here again, we are mainly interested in discussing the quantum limits of the strong cooling regime, where the frequency shift remains moderate but the effective damping allows efficient quench of Brownian fluctuations of the mechanical oscillator ($\Gamma_{eff} \gg \Gamma$). In this regime and for a peaked frequency response, the normalized position fluctuations reduce to:

$$\Delta X^2 \simeq \frac{\Gamma}{\Gamma_{eff}}\left[1+2n_i+\frac{\varphi_{nl}}{2}Q\widehat{S}_{f_{opt}}(\Omega=1)+\frac{\varphi_{nl}}{2}Q\widehat{S}_{f_{opt}}(\Omega=-1)\right] \qquad (27)$$

where $n_i=1/(\exp(\hbar\omega_0/k_BT)-1)$ is the average number of quanta in the environment at frequency $\omega_0$. Reaching the quantum ground state of the oscillator means reaching the limit $\Delta X^2=1$.

In case of negligible photothermal effects (taking the limit A=0) and in the strong cooling regime, Eq. 24 to 27 lead the usual conclusions concerning radiation pressure cooling: one needs to be in the good-cavity limit to cool the oscillator arbitrary close to its ground-state. We will not insist here on radiation-pressure cooling, which has already been discussed in details in several articles [10-13].

We focus on situations where photothermal pressure overcomes radiation pressure (βA≫1). This means that absorption at the back mirror is not negligible and generally implies that we are working in the "bad cavity limit", where $\omega_0 \ll \kappa$. Of course, "bad cavity" is just a naming here and the cavity finesse can in reality be large, provided that A and T are small enough.

In Fig. 2a, we have plotted the optical force spectral noise density for a set of parameters close to that considered in first section ($\varphi=1$, A=0.01, T=0.001, d=1, $\beta=10000$) and for a value of b=0.01 which places us deeply in the "bad cavity" limit. The comparison with Fig. 1a is instructive: the spectrum is now asymmetric in frequency and posses a radiation-pressure "bump" on top of the dominating photothermal contribution centered at null frequency. The overall noise is not simply the sum of photothermal and radiation-pressure contributions, it is a subtle interference between the two that will be discussed below.

Fig. 2b shows the normalized variance $\Delta X^2$ obtained from Eq. 27 as a function of the normalized detuning $\varphi=\Delta/\kappa$ and of the d parameter (d=$\omega_0\tau_{th}$), in the regime of strong cooling and in the limit of large $\varphi_{nl}$. As can be seen on the figure, the quantum ground-state is approached for a large set of values of $\varphi$ and d. A large value of detuning $\varphi$ for example reduces the influence of radiation pressure noise and is generally favorable to approach the ground-state. But the situation here offers other regimes of interest. For $\varphi=1$ for example, we observe that increasing the parameter d from 0 to 100 allows to reduce the position variance and also approach the ground-state. For values of d much larger than 100 here (that is well above $\beta$A), the variance increases again reflecting the fact that optomechanical damping of photothermal origin becomes less efficient and does not damp the response of the oscillator fluctuations to radiation pressure noise anymore. This increase of the position variance for large d was absent from the calculations of first section because we did not include radiation pressure effects in our first description. In reality, we see that the inclusions of photothermal and radiation pressure effects lead to the existence of an optimum value of d. Still our important conclusion remains: the ground-state can be reached by photothermal cavity cooling in the bad cavity limit and in a regime of moderate detuning where photons are more easily injected into the cavity by the pump field.

The minimum number of phonons in the oscillator that can be reached by cavity cooling is obtained directly from the optical force noise density spectrum using a detailed balance argument [10,27]:

$$n_{min} = \left[\frac{S_{f_{opt}}(-\omega_0)}{S_{f_{opt}}(\omega_0)} - 1\right]^{-1} \qquad (28)$$

Fig. 3 plots $n_{min}$ as a function of $d=\omega_0\tau_{th}$, for a detuning $\varphi=1$ (flank of the cavity resonance), in the "bad cavity" limit b=0.1 and for different values of βA. Occupation factors well below one are obtained here by an appropriate choice of the d parameter. More generally, the quantum ground-state is approached arbitrary closely in the "bad-cavity" regime by careful adjustment of β, A, T d and for moderate values of $\varphi\sim1$.

As discussed in [10,27], the asymmetry in the noise spectrum is responsible for a net exchange of energy between optical and mechanical resonators. The optomechanical damping is obtained directly from the force noise spectrum using the formula:

$$\Gamma_{opt} = \frac{x_{ZPF}^2}{\hbar^2}\left[S_{f_{opt}}(-\omega_0) - S_{f_{opt}}(\omega_0)\right] \qquad (29)$$

This formula can be seen as generalized Kubo formula [28]. Using the Eq. 24 in the "bad cavity" limit b<<1 and injecting the result into Eq. 29, we obtain an expression for the optomechanical damping on the "red-detuned" flank of the cavity resonance ($\varphi=1$) when the absorption dominates other losses in the cavity (T<<A):

$$\frac{\Gamma_{opt}}{\Gamma} = Q_m \beta \frac{\omega_0 \tau_{th}}{1+\omega_0^2 \tau_{th}^2} \frac{2}{c} P_{abs} \frac{8F}{\pi} \frac{1}{K} \qquad (30)$$

where we have made use of the following equivalence between normalization factors of the classical and quantum calculations: $1/\kappa\,(\sqrt{2}\hbar k\alpha/\tau)^2 = (2R/c)^2\,\hbar\omega_L P_{circ}\,F/\pi$ in the limit $R\sim 1$.

Eq. 30 is exactly the expression found classically in the first section (Eq. 7), showing a good level of consistency between classical and quantum calculations. But the quantum approach provides us now with an additional understanding: the photothermal damping results from an interference between radiation pressure and photothermal pressure noises. In absence of radiation pressure, the quantum noise spectrum of the optical force would have been symmetric like shown in Fig. 1a, precluding the possibility to cool the oscillator using photothermal back-action. It is really the synergy of both optical forces (photothermal and radiation pressure) which allows cooling. Classically it can be understood as the cavity providing an optical resonance in the interaction between photons and the mechanical oscillator and the photothermal effects providing a large force with the time lag necessary for dynamical back-action cooling.

## 4. Conclusions

We have presented a classical and a quantum treatment of the problem of cavity cooling of a mechanical oscillator by photothermal dynamical back-action. Both approaches are consistent and show that the quantum ground state of the oscillator can be reached using photothermal optomechanical cavity cooling, even in the "bad cavity" limit where the lifetime of photons in the cavity is smaller than the mechanical time period of the oscillator. This is in strong contrast to the case of cooling by radiation pressure and opens new experimental perspectives. Using photothermal cooling, the ground state could be reached without having to fulfill the "good cavity" condition and in situation of moderate cavity detuning where a large number of photons would be injected in the cavity.

Our calculations are valid for any semi-classical force that can be written in the form of Eq. 1 or Eq. 14. They could hence be adapted to other delayed photo-induced forces arising for example from radiometric pressure or photo-strictive effects. The choice of a simple exponential retardation function with a unique time-scale $\tau$ is motivated by our approach of neglecting the details of involved microscopic processes. As mentioned above, for what concerns the photothermal force, this choice is validated by several experimental studies [14,16-18]. A more a complex behavior of the force, represented for example by several time-scales, would only modify quantitatively the force noise spectrum and not affect qualitatively our conclusions.

Acknowledgments:

We thank Florian Marquardt and Simone De Liberato for stimulating discussions and the C-Nano Ile de France for financial support.

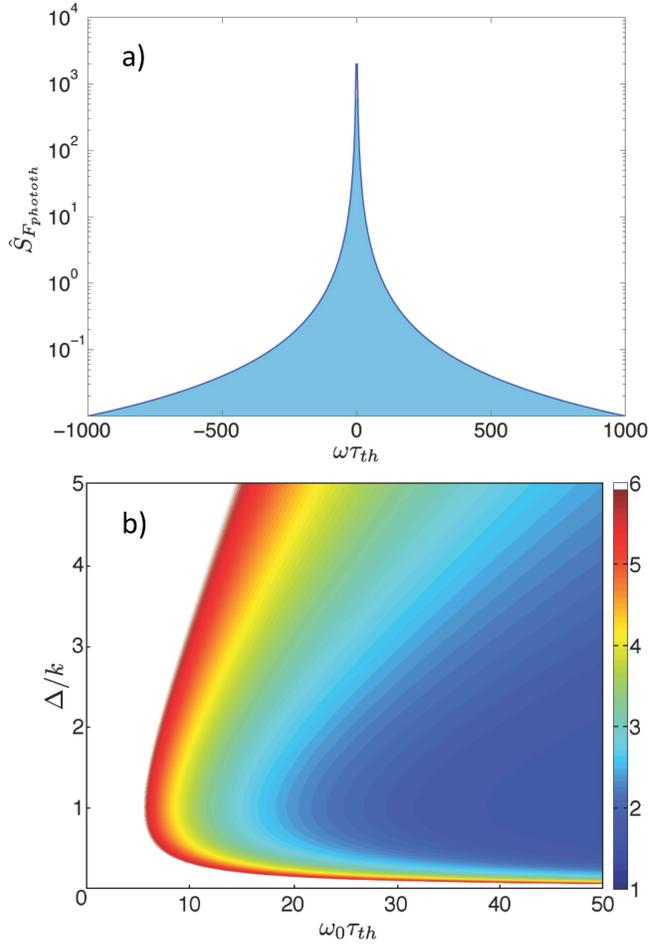

Fig. 1. Classical calculation without radiation pressure. (a) Normalized noise spectrum of the photothermal force for A=0.01, β=$10^4$ and T=0.001. (b) Normalized fluctuation temperature $2m\omega_0^2\langle x^2\rangle/(\hbar\omega_0)$ of the mechanical oscillator as a function of the normalized detuning $\Delta/\kappa$ and $\omega_0\tau_{th}$.

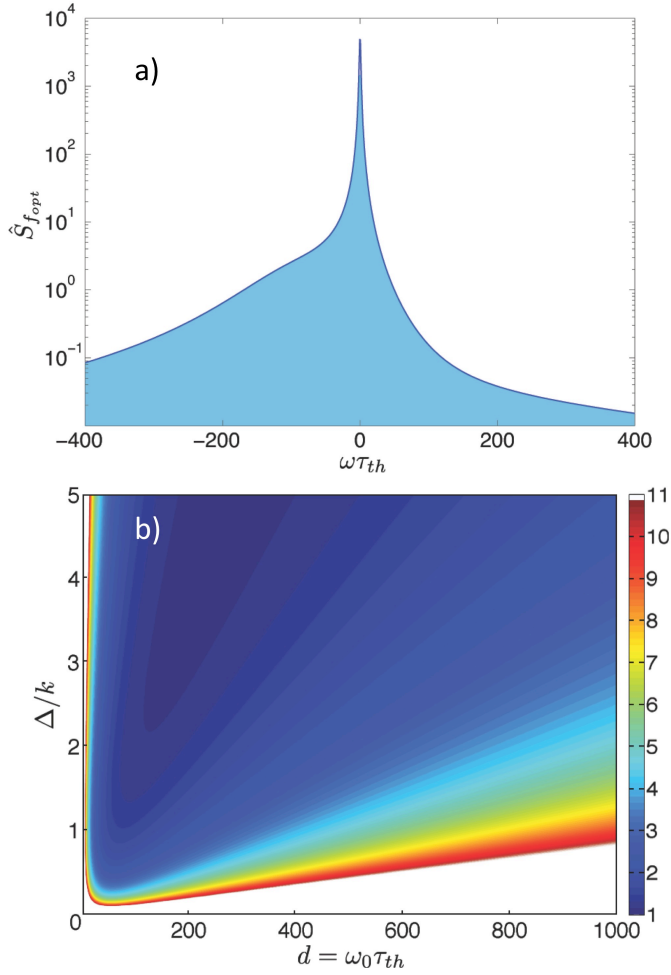

Fig. 2. Quantum calculation including both photothermal and radiation pressure. (a) Normalized noise spectrum of the complete optical force, including both radiation and photothermal pressures, for A=0.01, β=$10^4$, T=0.001, φ=1, c=1 and in a bad cavity situation b=0.01. (b) Normalized variance $X^2$ as a function of the normalized detuning Δ/κ and of the parameter d=$\omega_0\tau_{th}$.

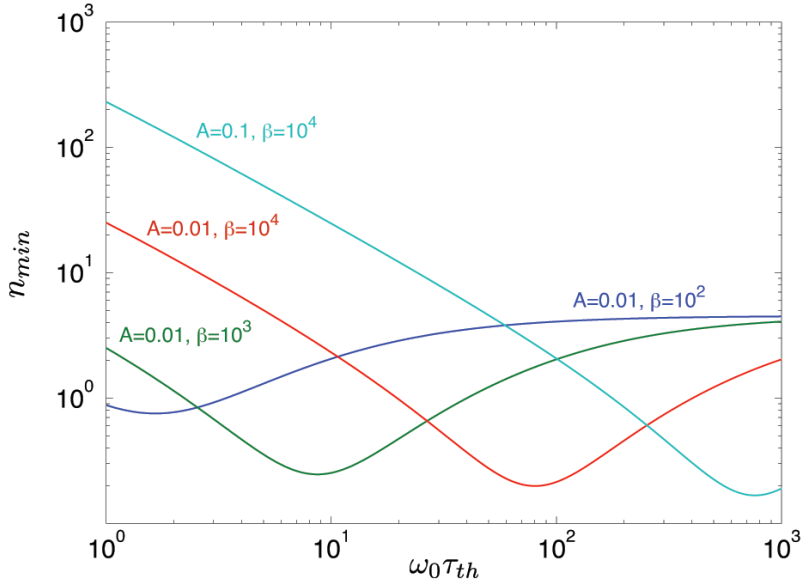

Fig. 3. Minimum phonon occupation of the mechanical oscillator as a function of $c=\omega_0\tau_{th}$, for a detuning $\varphi=1$ and for various values of the parameters $\beta$ and $A$.